\documentclass[prd,superscriptaddress,nofootinbib,amsmath,amssymb,aps,10pt]{revtex4}

\usepackage{bm}
\usepackage{amsfonts}
\usepackage{latexsym}
\usepackage[latin1]{inputenc}
\usepackage{graphicx}
\usepackage{amsmath}
\usepackage{palatino}
\usepackage{mathpazo}
\usepackage[outdir=./]{epstopdf}

\usepackage{booktabs}
\usepackage{dcolumn}

\def\jnl@style{\it}
\def\aaref@jnl#1{{\jnl@style#1}}

\def\aaref@jnl#1{{\jnl@style#1}}

\def\aj{\aaref@jnl{AJ}}                   % Astronomical Journal
\def\apj{\aaref@jnl{ApJ}}                 % Astrophysical Journal
\def\apjl{\aaref@jnl{ApJ}}                % Astrophysical Journal, Letters
\def\apjs{\aaref@jnl{ApJS}}               % Astrophysical Journal, Supplement
\def\apss{\aaref@jnl{Ap\&SS}}             % Astrophysics and Space Science
\def\aap{\aaref@jnl{A\&A}}                % Astronomy and Astrophysics
\def\aapr{\aaref@jnl{A\&A~Rev.}}          % Astronomy and Astrophysics Reviews
\def\aaps{\aaref@jnl{A\&AS}}              % Astronomy and Astrophysics, Supplement
\def\mnras{\aaref@jnl{Mon.~Not.~Roy.~Astron.~Soc.}}             % Monthly Notices of the RAS
\def\prd{\aaref@jnl{Phys.~Rev.~D}}        % Physical Review D
\def\prc{\aaref@jnl{Phys.~Rev.~C}}  % Physical Review C
\def\prl{\aaref@jnl{Phys.~Rev.~Lett.}}    % Physical Review Letters
\def\qjras{\aaref@jnl{QJRAS}}             % Quarterly Journal of the RAS
\def\skytel{\aaref@jnl{S\&T}}             % Sky and Telescope
\def\ssr{\aaref@jnl{Space~Sci.~Rev.}}     % Space Science Reviews
\def\zap{\aaref@jnl{ZAp}}                 % Zeitschrift fuer Astrophysik
\def\nat{\aaref@jnl{Nature}}              % Nature
\def\aplett{\aaref@jnl{Astrophys.~Lett.}} % Astrophysics Letters
\def\apspr{\aaref@jnl{Astrophys.~Space~Phys.~Res.}} % Astrophysics Space Physics Research
\def\physrep{\aaref@jnl{Phys.~Rep.}}      % Physics Reports
\def\physscr{\aaref@jnl{Phys.~Scr}}       % Physica Scripta
\def\commat{\aaref@jnl{Comm.~Math.~Phys.}}              % Communications in Mathematical Physics
\def\science{\aaref@jnl{Science}}               % Science
\def\cqg{\aaref@jnl{Classical Quant.~Grav.}}            % Classical and Quantum Gravity
\def\jpcs{\aaref@jnl{JPCS}}                                     % Journal of Physics Conference Series
\def\ijmpd{\aaref@jnl{Int.~J.~Mod.~Phys.~D}}                    % International Journal of Modern Physics D
\def\grg{\aaref@jnl{Gen.~Relat.~Gravit.}}               % General Relativity and Gravitation
\def\rpp{\aaref@jnl{Rep.~Prog.~Phys.}}          % Reports on Progress in Physics
\def\npa{\aaref@jnl{Nucl.~Phys.~A}}        % Nuclear Physics A
\def\lrr{\aaref@jnl{Living Rev.~Rel.}}                   % Living reviews in relativity
\def\jcap{\aaref@jnl{J.~Cosmology Astropart.~Phys.}}    % Journal of cosmology and astroparticle physics
\def\rmp{\aaref@jnl{Rev.~Mod.~Phys.}}   %Reviews of modern physics
\def\rmp{\araa@jnl{Annu.~Rev.~Astron.~Astrophys.}}   %Reviews of modern physics

\allowdisplaybreaks[1]
\begin{document}

\title{Differentially rotating neutron stars in scalar-tensor theories of gravity}

\author{Daniela D. Doneva}
\email{daniela.doneva@uni-tuebingen.de}
\affiliation{Theoretical Astrophysics, Eberhard-Karls University
	of T\"ubingen, T\"ubingen 72076, Germany}
\affiliation{INRNE - Bulgarian Academy of Sciences, 1784  Sofia, Bulgaria}

\author{Stoytcho S. Yazadjiev}
\email{yazad@phys.uni-sofia.bg}
\affiliation{Department
	of Theoretical Physics, Faculty of Physics, Sofia University, Sofia
	1164, Bulgaria}
\affiliation{Theoretical Astrophysics, Eberhard-Karls University
	of T\"ubingen, T\"ubingen 72076, Germany}
\affiliation{Institute of Mathematics and Informatics, 	Bulgarian Academy of Sciences, 	Acad. G. Bonchev St. 8, Sofia 1113, Bulgaria}

\author{Nikolaos Stergioulas}
\email{niksterg@auth.gr}
\affiliation{Department of Physics, Aristotle University of Thessaloniki, Thessaloniki 54124, Greece}

\author{Kostas~D.~Kokkotas}
\email{kostas.kokkotas@uni-tuebingen.de}
\affiliation{Theoretical Astrophysics, Eberhard-Karls University
	of T\"ubingen, T\"ubingen 72076, Germany}

\begin{abstract}
We present the first numerical models of differentially rotating stars in alternative theories of gravity. We chose a particular class of scalar-tensor theories of gravity that is indistinguishable from GR in the weak field regime but can lead to significant deviations when strong fields are considered.
We show that the maximum mass that a differentially rotating neutron star can sustain increases significantly for scalarized solutions and such stars can reach larger angular momenta. In addition, the presence of a nontrivial scalar field has the effect of increasing the required axis ratio for reaching a given value of angular momentum, when compared to a corresponding model of same rest mass in general relativity. We find that the scalar field also makes rapidly rotating models less quasi-toroidal than their general-relativistic counterparts. For large values of the angular momentum and values of the coupling parameter that are in agreement with the observations, we find a second turning point for scalarized models along constant angular momentum sequences, which could have interesting implications for the stability of remnants created in a binary neutron star merger.   
\end{abstract}

\maketitle

\section{Introduction}
The gravitational wave observations of merging neutron stars open new horizons towards testing our fundamental understanding of the laws of nature, such as the behavior of matter at extreme densities and the strong field regime of gravity. Such mergers are being studied with fully nonlinear three-dimensional, general-relativistic simulations, which are computationallly very demanding (see \cite{Baiotti:2016qnr,Paschalidis2017,2018PhRvL.120v1101D} and references therein). Very valuable conclusions on the fate of binary neutron star mergers can also be reached by examining equilibrium sequences of rotating neutron stars (see e.g. \cite{Breu2016,Bauswein2017a,Weih2018,Bozzola2018}). The remnants formed after a binary neutron star merger have strong differential rotation, which can temporarily delay the collapse to a black hole. In the case of neutron star formation through core collapse, even though the standard scenario is a slowly rotating nascent neutron star, a rapidly and differentially rotating neutron star cannot be excluded for some progenitors.

Methods for studying equilibrium models of differentially rotating neutron stars in general relativity (GR) are well established (see  \cite{Friedman2013,Paschalidis2017} and reference therein). In contrast to the uniformly rotating case, where a model is defined uniquely by three parameters, in the case of differential rotation different types of equilibrium models can exist \cite{2004MNRAS.352.1089S, 2006PhRvL..96p1101Z, Ansorg2009}, such as quasi-toroidal configurations and additional types. In the case of binary neutron star mergers, simulations show that the remant is still quasi-spherical, but with a rotational profile such that the maximum angular velocity appears off-center (see, e.g. \cite{Kastaun:2014fna, Kastaun:2016yaf,Hanauske:2016gia}).

To date, models of differentially rotating neutron stars have been constructed only in GR. Even though Einstein's theory of gravity is very well tested in the weak field regime (mainly with the solar system experiments), constraints in the strong field regime (such as through gravitational waves from the coalescence and merger of compact objects) still have large error bars (e.g. \cite{PhysRevLett.120.031104, PhysRevLett.116.221101}). Further tests using gravitational waves in the near future, could narrow the constraints in the strong-field regime. For binary neutron star mergers, the uncertainties in the equation of state (EOS) and in the correct theory of gravity need to be disentangled. In the present work we take a step in this direction by examining, for the first time, differentially rotating equilibrium models of neutron stars in well-known alternative theory of gravity -- the scalar-tensor theory. Our equilibrium sequences could be used to extend the universal (EOS-independent) relations found in GR \cite{Breu2016,Bauswein2017a,Weih2018,Bozzola2018} and thus obtain some insight regarding the outcome neutron star mergers, without performing nonlinear simulations.

Neutron star models in scalar-tensor theories became particularly popular after the discovery by  Damour and Esposito-Farese \cite{Damour1993} that in certain classes of scalar field coupling functions a nontrivial structure of the solutions is present. That is, in addition to the GR neutron stars that are always solutions of the field equations, additional branches of neutron stars with nontrivial scalar field exist for certain range of the parameters. Moreover, these theories are perturbatively equivalent to GR in the weak field regime so that they satisfy all solar system experiments. Most importantly, these scalarized solutions are energetically favoured over the corresponding GR solutions \cite{Damour1993,Harada1997,Harada1998}.
Slowly rotating neutron stars were constructed in \cite{Damour1996,Sotani2012,Pani2014a,Silva2015} and the case of rapid rotation was examined in \cite{Doneva2013,Doneva2016}, where it was shown that neutron star models with high rotation rates can deviate significantly from the corresponding GR models (in contrast to case of nonrotating or slowly rotating models). Currently, the binary pulsar experiments constrain heavily the values of the free parameters of the theory \cite{Demorest10,Antoniadis13,Shao2017} which leaves very little space for deviations from GR in the nonrotating case, while the regime of rapid rotating opens new prospects in testing scalar-tensor theories. We note that in the case of a binary neutron star merger, scalarization can occur dynamically during inspiral \cite{Barausse2013,Shibata2014}.

The structure of the paper is as follows: In Section \ref{sec:sct} we review briefly the theory behind constructing equilibrium differentially rotating neutron stars in scalar-tensor theories. The numerical results are presented in Section \ref{sec:resuts}. Conclusions are presented in Section \ref{sec:conc}.

\section{Rotating stars in scalar-tensor theories of gravity}
\label{sec:sct}
The general form of the Einstein frame action in scalar-tensor theories (STT) is   
\begin{eqnarray}
S= {1\over 16\pi G_{*}}\int d^4x \sqrt{-g} \left(R -
2g^{\mu\nu}\partial_{\mu}\varphi \partial_{\nu}\varphi -
4V(\varphi)\right)+ S_{m}[\Psi_{m}; {\cal A}^{2}(\varphi)g_{\mu\nu}],
\end{eqnarray}
where $R$ is the Ricci  scalar with respect to the Einstein frame metric $g_{\mu\nu}$, $V(\varphi)$ is the scalar field potential and $ {\cal A}^{2}(\varphi)$ is the Einstein frame coupling function between the matter and the scalar field that appears in the action of the matter $S_{m}$. The matter fields are collectively denoted by $\Psi_{m}$.  After varying the action one can obtain the following field equations
\begin{eqnarray} 
R_{\mu\nu} - {1\over 2}g_{\mu\nu}R &=& 8\pi G_{*} T_{\mu\nu}
+ 2\partial_{\mu}\varphi \partial_{\nu}\varphi   -
g_{\mu\nu}g^{\alpha\beta}\partial_{\alpha}\varphi
\partial_{\beta}\varphi -2V(\varphi)g_{\mu\nu}  \,\,\, , \label{EFFE1}\\
\nabla^{\mu}\nabla_{\mu}\varphi &=& - 4\pi G_{*} k(\varphi)T
+ {dV(\varphi)\over d\varphi} , \label{EFFE}
\end{eqnarray}
where $k(\varphi)$ is defined as
\begin{equation}
k(\varphi)= \frac{d\ln({\cal  A}(\varphi))} {d\varphi}.
\end{equation}

In the present paper we will consider a class of STT with zero scalar field potential $V(\varphi)=0$ and coupling function 
\begin{equation}
k(\varphi) = \beta \varphi,
\end{equation}
where $\beta$ is a constant. This case is particularly interesting, since the resulting STT is perturbatively equivalent to GR in the weak field regime while, nonlinear effects, such as scalarization of neutron stars, can be observed for strong fields.

We consider equilibrium configurations, in which the matter and scalar field (and hence the spacetime) are stationary and axisymmetric. In this case, the metric can take the following general form:
\begin{eqnarray}
&&ds^2 = -e^{\gamma+\sigma} dt^2 + e^{\gamma-\sigma} r^2
\sin^2\theta (d\phi - \omega dt)^2 + e^{2\alpha}(dr^2 + r^2
d\theta^2),
\end{eqnarray}
where all metric functions depend only on  $r$ and $\theta$. The explicit form of the dimensionally reduced field equations are given in \cite{Doneva2013,Yazadjiev2015,Doneva2016}. In this section, we will discuss in detail only the equation for hydrostationary equilibrium, since it depends directly on the chosen rotation law. This equation can be derived from the conservation of the energy-momentum tensor that takes the following form in the Einstein frame
\begin{eqnarray}
\nabla_{\mu}T^{\mu}{}_{\nu} = k(\varphi)T\partial_{\nu}\varphi .
\end{eqnarray}
For a perfect fluid
\begin{eqnarray}
T_{\mu\nu}= (\varepsilon + p)u_{\mu} u_{\nu} + pg_{\mu\nu},
\end{eqnarray}
where $p$ and $\varepsilon$ are the Einstein frame pressure and energy density of the fluid. One can easily show that the Einstein frame fluid four velocity takes the form
\begin{equation}
u^\mu = \frac{e^{-(\sigma + \gamma)/2}}{\sqrt{1-v^2}}
[1,0,0,\Omega], \label{eq:four_velocity}
\end{equation}
where the angular velocity is defined as 
\begin{equation}
\Omega=\frac{u^{\phi}}{u^{t}}, \label{eq:Omega_definition}
\end{equation}
and the proper velocity $v$ of the fluid is given by
\begin{equation}
v = (\Omega - \omega) r \sin \theta e^{-\sigma}.
\end{equation}
For an uniformly rotating star, $\Omega$ is a constant throughout the star, while for differential rotation $\Omega=\Omega(r,\theta)$.

Up to this point, we defined quantities in the Einstein frame for convienience in the calculations, which differs from the physical frame (Jordan frame) by a conformal transformation of the metric plus a redefinition of the scalar field. After completing the calculations in the Einstein frame, quantities of interest will be reported in the physical Jordan frame. Detailed relations between the two frames, especially in the case of rapidly rotating neutron stars, can be found in \cite{Doneva2013}. Here, we will only mention some of the basic relations that are directly relevant for the results presented in Section \ref{sec:resuts}.

The energy-momentum tensor, the energy density, the pressure and the four velocity transform between the two frames as
\begin{eqnarray}\label{DPTEJF}
T_{\mu\nu} &=& {\cal A}^2(\varphi){\tilde T}_{\mu\nu} \nonumber \\
\varepsilon &=&{\cal A}^4(\varphi){\tilde\varepsilon}, \nonumber \\
p&=&{\cal A}^4(\varphi){\tilde p},  \\
u_{\mu}&=& {\cal A}^{-1}(\varphi){\tilde u}_{\mu} . \nonumber
\end{eqnarray}
where the Jordan frame quantities are denoted with a tilde. $\Omega$ and $v$ remain the same in both frames. 

With the above assumptions, the hydrostationary equilibrium is described by
\begin{eqnarray}
\frac{\partial_i{\tilde p}}{{\tilde \varepsilon} + \tilde{p}} -
\left[\partial_i(\ln \, u^t) - u^t u_\phi \partial_i \Omega -
k(\varphi) \partial_i \varphi\right]=0 \label{eq:Hydrostatic_Equil}.
\end{eqnarray}
In Eq. (\ref{eq:Hydrostatic_Equil}) we have deliberately used the Jordan frame pressure and energy density. The reason is that the field equations and the equation for hydrostationary equilibrium have to be supplemented with an equation of state for high-density matter, which is given in terms of ${\tilde \varepsilon}$ and ${\tilde p}$.

For a barotropic EOS, where ${\tilde p}={\tilde p}({\tilde \varepsilon})$, the integrability condition of equation \eqref{eq:Hydrostatic_Equil} leads to the requirement that the product $u^t u_\phi$ is a function only of the angular  velocity $\Omega$, i.e. $u^t u_\phi =F(\Omega)$. In the present paper, we will work with one of the most standard and widely used choices for $F(\Omega)$, namely
\begin{equation}
F(\Omega) = A_{\rm diff}^2 (\Omega_c - \Omega),
\end{equation}
where $A_{\rm diff}$ is a parameter and $\Omega_c$ is the central value of the angular velocity \cite{Komatsu1989,Komatsu1989a}. Of course other choices of $F(\Omega)$ are possible but since this is the first study of differentially rotating neutron stars in alternative theories of gravity we will limit ourselves to the classical case and leave the exploration of other functions $F(\Omega)$, especially the ones relevant for neutron star mergers \cite{Kastaun:2014fna, Kastaun:2016yaf,Hanauske:2016gia,Paschalidis2017,Bauswein2017a,Bozzola2018}, for a future study.  

Let us comment on the relation between the Jordan and the Einstein frame mass, radius and angular momentum of the star. The tensor mass of the neutron stars is by definition the ADM mass in the Einstein frame that can be calculated either via an integral throughout the star or from the asymptotics of the Einstein frame metric \cite{Stergioulas95,Friedman2013,Doneva2013}. Moreover, for the particular coupling functions  ${\cal A}^{2}(\varphi)$ considered in the present paper, this mass coincides with the Jordan frame one. The angular momentum is by definition the same in the two frames. What differs between the two frames, though, is the expression for the radius of the star. Thus the Jordan frame stellar radius is
\begin{equation}
{\tilde R}_e = {\cal A}(\varphi)\; r \;
e^{(\gamma-\sigma)/2}|_{r=r_{e},\theta=\pi/2},
\end{equation}
where $r_e$ is defined to be the Einstein frame coordinate equatorial radius of the star (corresponding to the location where the pressure vanishes ${\tilde p}(r_{e},\theta=\pi/2)=0$).

\section{Main results}
\label{sec:resuts}

\subsection{Constant angular momentum sequences}
In the present paper we focus on a single representative equation of state, namely the APR4 \cite{AkmalPR}, since our goal is to study the effect of scalarization on differentially rotating neutron stars, rather that making an extensive study of the parameter space. The numerical solution of the reduced field equations is performed using a new version of the {\tt RNS} code 
\cite{Stergioulas95} with differential rotation \cite{2004MNRAS.352.1089S} extended to STT \cite{Doneva2013,Yazadjiev2015,Doneva2016}. This code is based on the KEH method \cite{Komatsu1989,Komatsu1989a} with the modifications introduced in \cite{Cook1992,Cook1994} and has proved to be very reliable and robust for rapidly rotating neutron star models in GR, both uniformly and differentially rotating. Here we report results for ${\hat A} = A_{\rm diff}/r_e = 1.225$ that was used in previous studies of differentially rotating neutron stars in GR (in the context of the stability of post-merger remnants \cite{Bauswein2017a}).

\begin{figure}[]
	\centering
	\includegraphics[width=0.45\textwidth]{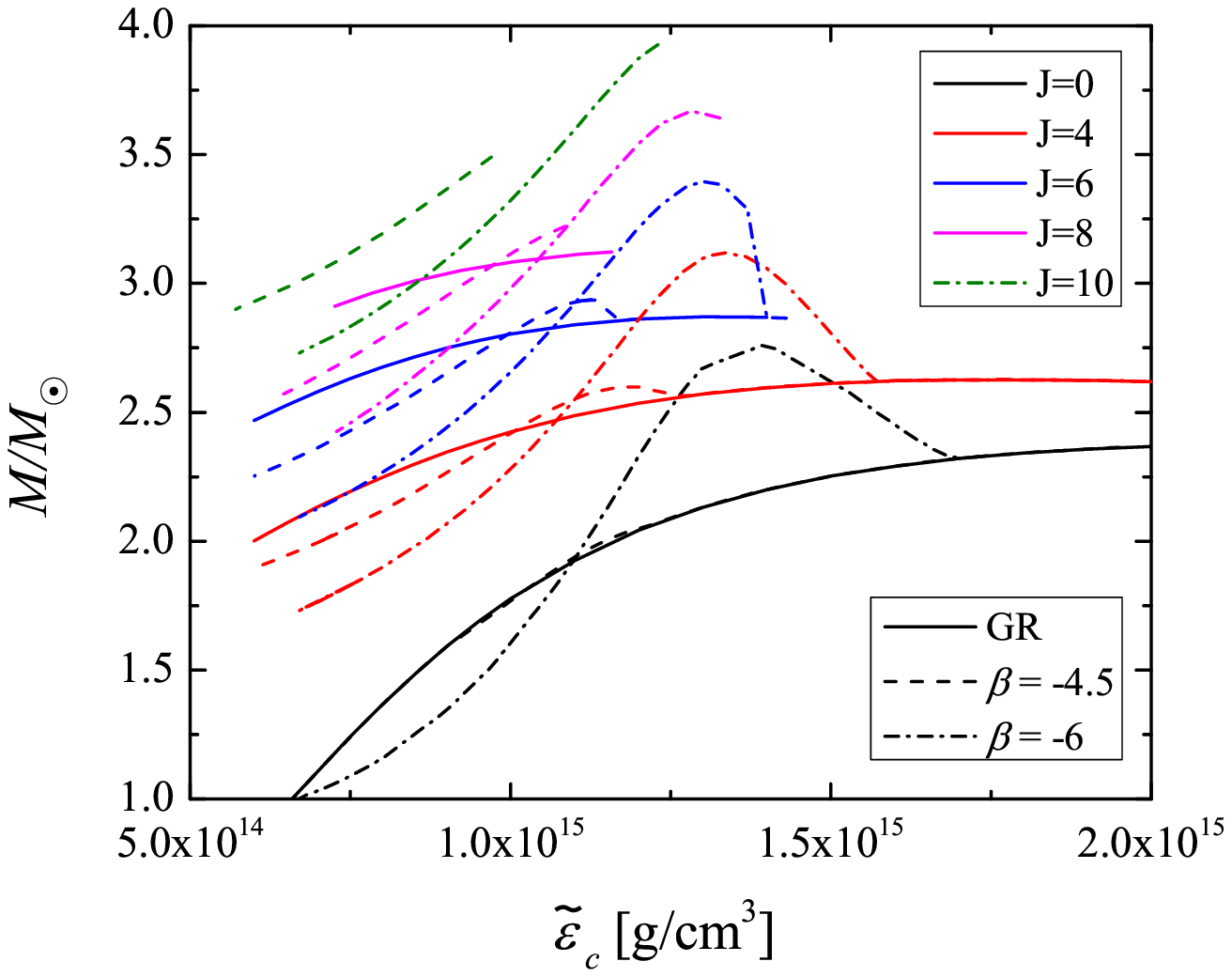}
	\includegraphics[width=0.45\textwidth]{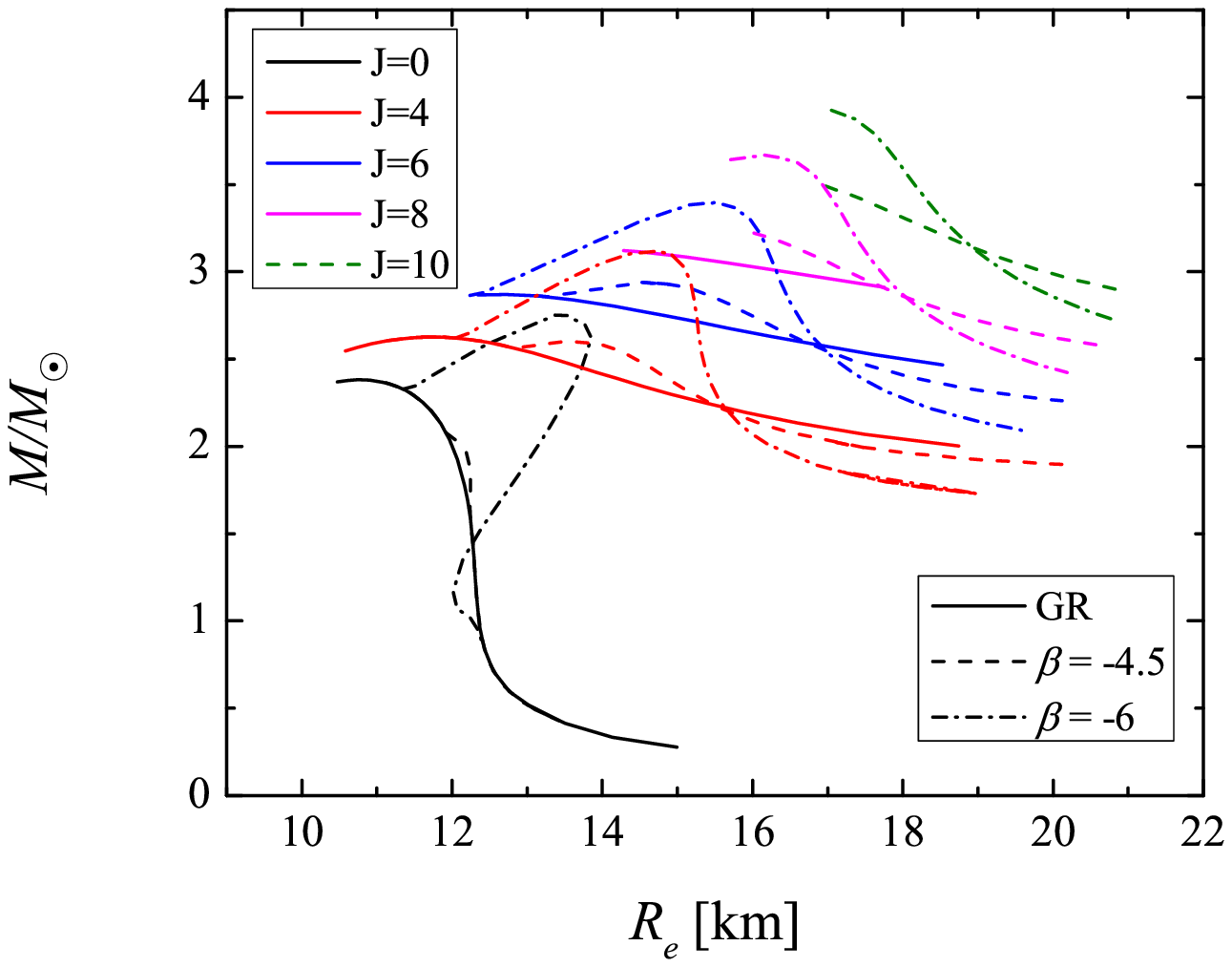}
	\caption{\textit{Left panel:} The mass as a function of the central energy density for differentially rotating sequences of neutron stars with constant angular momentum ($J$ is given in geometrical units). All the models have fixed ${\hat A} = A_{\rm diff}/r_e = 1.225$. Sequences are shown for $\beta = 0$ (GR), $\beta=-4.5$ and $\beta=-6$. \textit{Right panel:} Same as left panel, but as a function of radius.}
	\label{Fig:M_epsc_R}
\end{figure}

In Fig. \ref{Fig:M_epsc_R} we plot sequences with constant angular momentum $J$ (in geometrical units), both in the GR case and for STT with different values of $\beta$. In the left panel the mass $M$ as a function of the central energy density ${\tilde \varepsilon}_c$ is shown, while in the right panel the mass as a function of the neutron star radius $R_e$ is plotted. We have chosen to present constant $J$ sequences since their turning points in an $M({\tilde \varepsilon}_c)$ plot are related to the change of quasi-radial stability. There is no turning-point theorem for the onset of instability for differentially rotating stars, but simulations have shown that the actual dynamical instability is relatively close to the turning points \cite{Weih2018}. The model with smallest central energy density for each sequence with $J>0$ is at the mass-shedding limit.  The sequences are terminated at a central energy density only somewhat higher than the turning point. Notice that for high angular momentum ($J>6$) the sequences terminate at a central energy density somewhat smaller than the actual turning point. The reason for this is the following. At axis ratios smaller than 0.5, models can become quasi-toroidal (more generally, several types of differentially rotating solutions can exist \cite{Ansorg2003}) and that is why in some parts of the parameter space exactly close to the turning point, the numerical method in {\tt RNS} may not converge to a unique solution. This problem could be circumvented using a different formulation, as in \cite{Ansorg2003}. 

\begin{figure}[]
	\centering
	\includegraphics[width=0.45\textwidth]{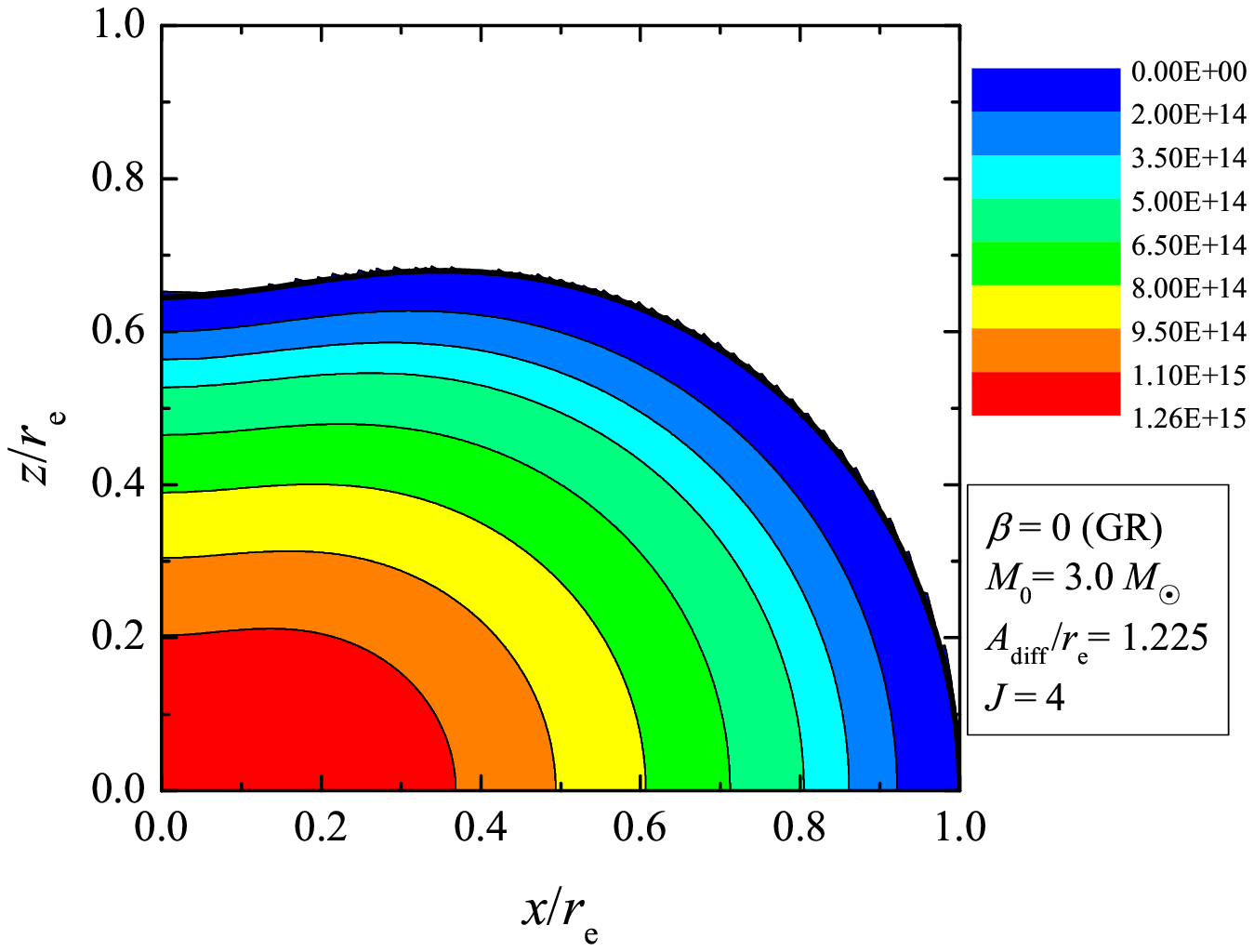}
	\includegraphics[width=0.45\textwidth]{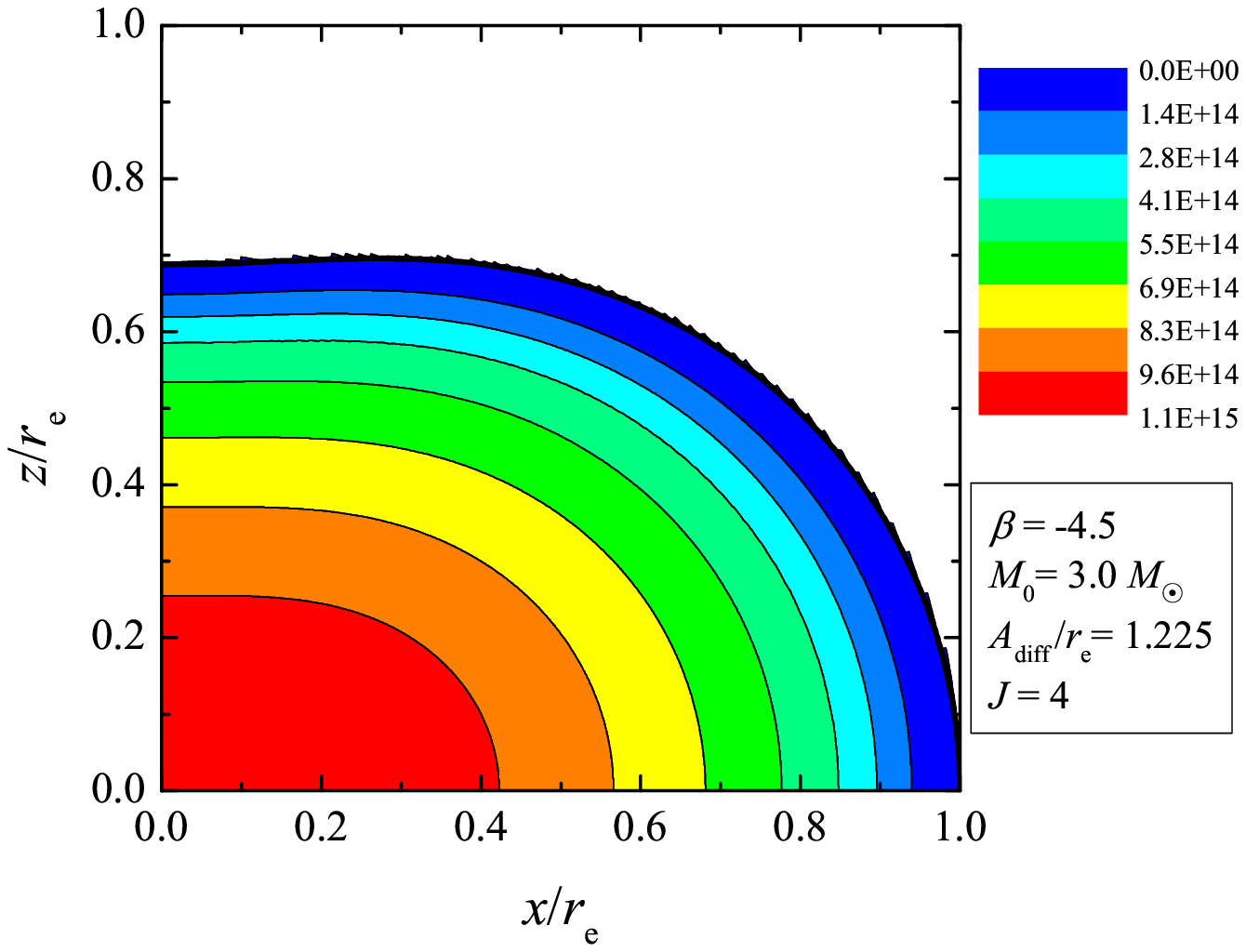}
	\caption{Contour plots of the neutron star energy density for a model with ${\hat A} = A_{\rm diff}/r_e = 1.225$, $J=4$ (in geometrical units)
	and baryon mass $M_0=3.0M_\odot$. \textit{Left panel:}  $\beta=0$ (GR). \textit{Right panel:} $\beta=-4.5$, notice that the scalarized model has a somewhat larger axis ratio and the core is less quasi-toroidal.}
	\label{Fig:Model_J4eps}
\end{figure}

\begin{figure}[]
	\centering
	\includegraphics[width=0.45\textwidth]{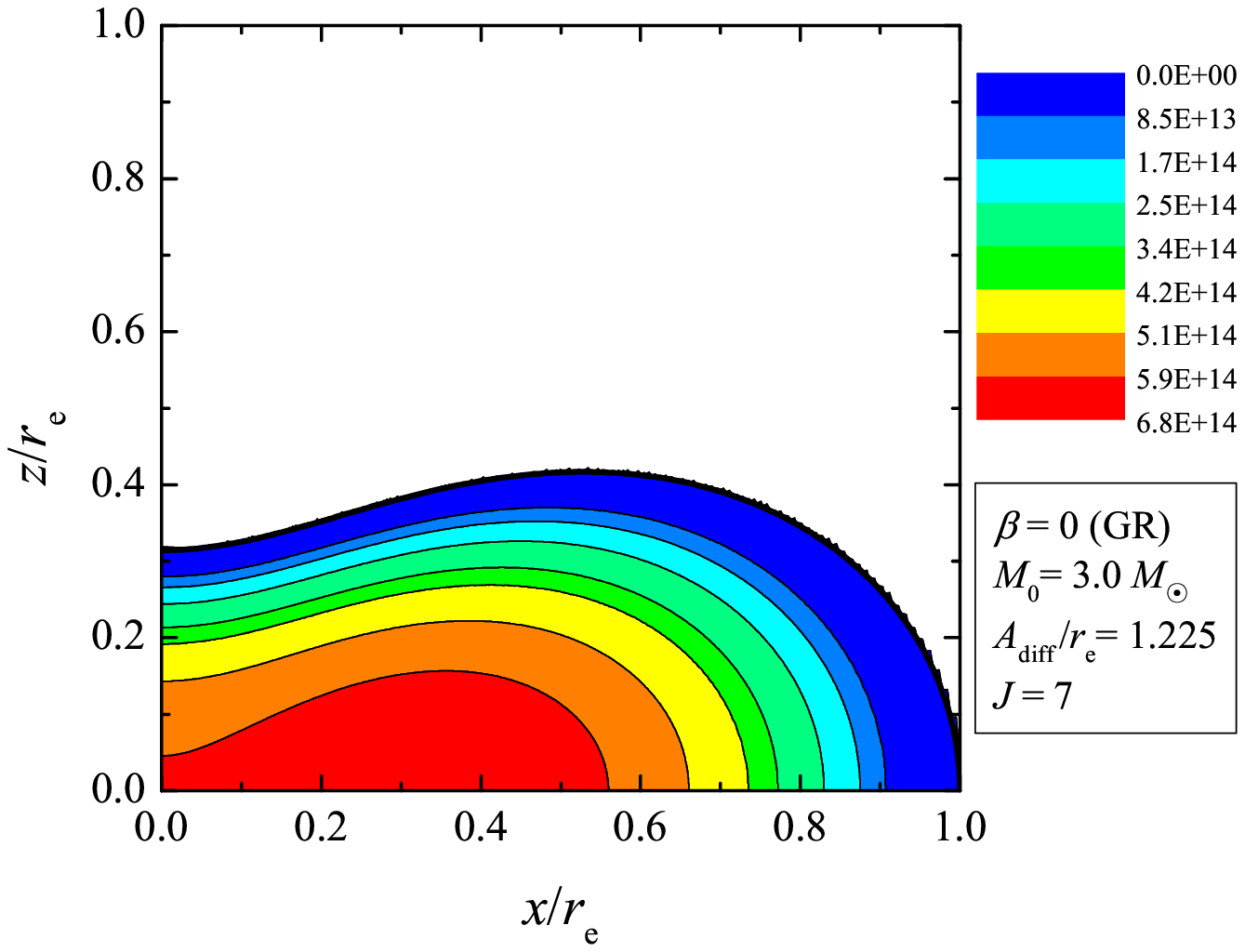}
	\includegraphics[width=0.45\textwidth]{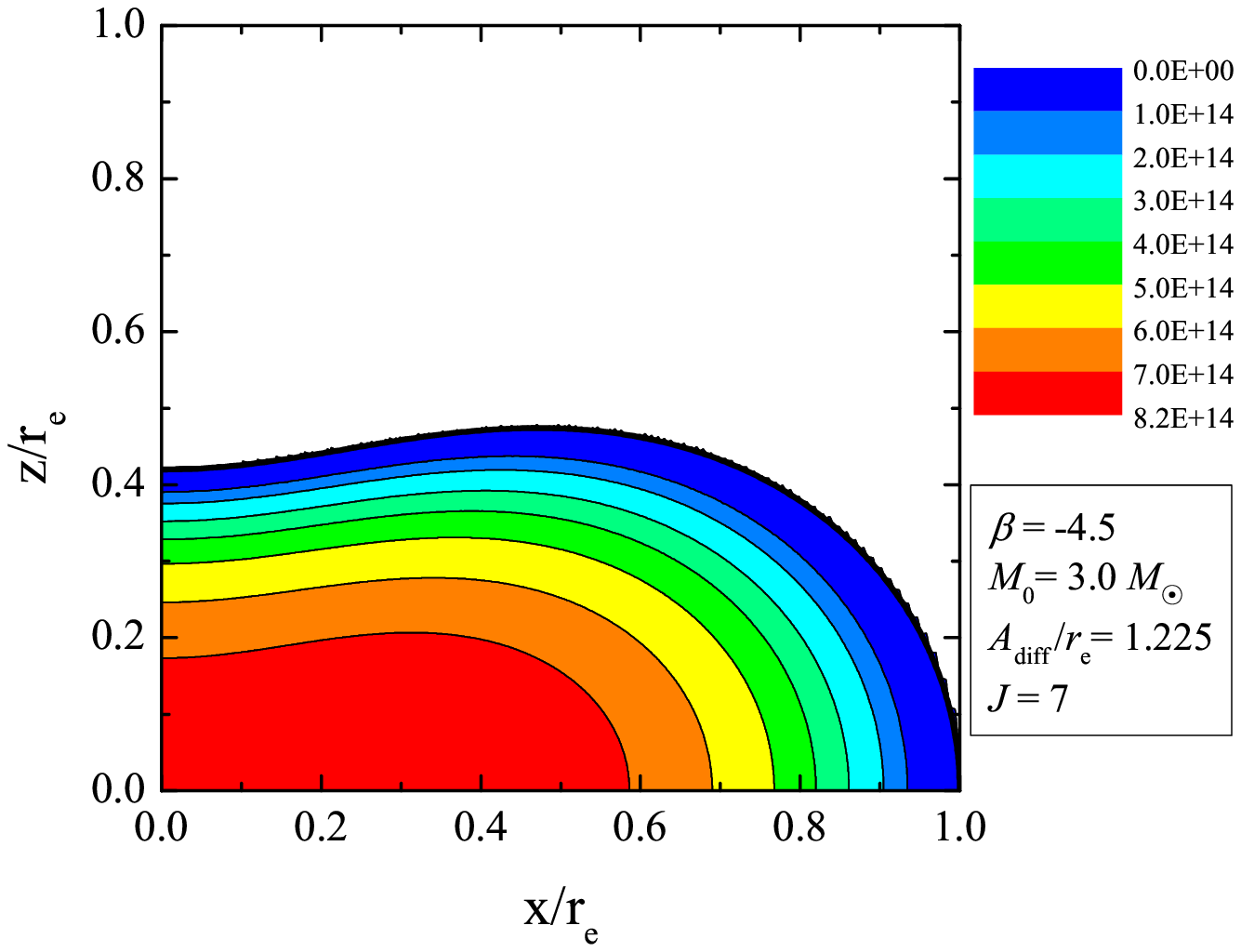}
	\caption{Same as Fig. \ref{Fig:Model_J4eps}, but for $J=7$.}
	\label{Fig:Model_J7eps}
\end{figure}

As $J$ increases, the central energy density of the model at the mass-shedding limit increases, whereas the central energy density of the model close to the turning point decreases. For high $J$ the range of central energy densities for which models exist thus becomes shorter and for $J\gtrsim 9$ there is no equilibrium sequence in GR, for the chosen rotation law (see Fig. \ref{Fig:M_epsc_R})  (notice that uniformly rotating models are limited to $J\lesssim 4$ in geometrical units). 

We find that scalarized differentially rotating solutions can reach larger values of $J$, depending on the chosen value of the parameter $\beta$. In Fig. \ref{Fig:M_epsc_R} we show results up to $J=10$, but the particular threshold value of $J$ above which no solutions can be obtained would depend on the parameter $\beta$. We expect that the existence of models with higher $J$ in scaralized solutions is related to the modification of the matter distribution inside the star, when a scalar field is present. 
 
In  Fig. \ref{Fig:M_epsc_R} results for two values of $\beta$ are presented, $\beta=-4.5$ and $\beta=-6$, but if we consider massless STT only $\beta=-4.5$ is within the observational limits implied by \cite{Antoniadis13,Demorest10}. The examination of models with  $\beta=-6$ can be justified on the following grounds: First, since this is the first study of differentially rotating neutron stars in STT, we want to study in detail the effect of varying $\beta$ on the results. Second, considering $\beta<-4.5$ can be justified also if the scalar field has nonzero mass. In this case the scalar field will be confined within its Compton wavelength that is directly related to its mass. Thus, for large enough values of the mass the emission of scalar gravitational radiation could be suppressed, which would reconcile the theory with the binary pulsar observations for a larger range of $\beta$ \cite{Popchev2015,Ramazanouglu2016,Yazadjiev2016}. The massless case for different $\beta$ shown in Fig. \ref{Fig:M_epsc_R} is actually the upper limit for the deviations from GR in the massive case for the same values of $\beta$, which justifies our choice of examining  scalarized models with $\beta<-4.5$.

Even if we limit our study to the $\beta=-4.5$ case, an interesting observation can be made. While the scalarized neutron stars with $\beta=-4.5$ are almost indistinguishable from their GR counterparts for $J=0$, a clear difference appears for higher $J$. In addition, the scalarized neutron stars reach larger values of $J$, when compared to the GR case, as discussed above. The maximum mass that a differentially rotating neutron star can sustain therefore is higher for scalarized model than for GR models (with a strong dependence on the particular value of $\beta$).  Notice that in the present paper we compare differentially rotating sequences of scaralized models with constant $J$ to their GR counterparts, whereas in \cite{Doneva2013,Doneva2016} sequences at the mass-shedding limit were considered (this explains the somewhat smaller deviations from GR, for same $\beta$, in the current results, with respect to those observerd in \cite{Doneva2013,Doneva2016}).

In  Fig. \ref{Fig:M_epsc_R}  we also make the following observation: scalarized solutions with sufficient angular momentum (e.g. $J=4$, or $J=6$) posses two turning points along a constant-$J$ sequence for values of the parameter $\beta$ that are in agreement with the observations in the massless STT case, i.e. $\beta>-4.5$, while only one turning point appears for smaller values of $J$. The turning point at higher central density is the same as in GR. But, a second turning-point appears at somewhat lower central density. Such phenomenon, i.e. the appearance of two turning points, is also observed for more negative values of $\beta$ even in the nonrotating case, but if we consider massless STT, they are not within the observational limits. Since in GR turning points are associated with quasi-radial stability and the scalarized solution is energetically favoured over the GR solution (in the region where scalarized solutions appear) it will be important to study the dynamical properties of models along such constant-$J$ sequences, in order to determine whether an unstable region exists between the two turning points (which would have important astrophysical consequences for the stability of binary neutron star merger remnants).

\begin{figure}[]
	\centering
	\includegraphics[width=0.45\textwidth]{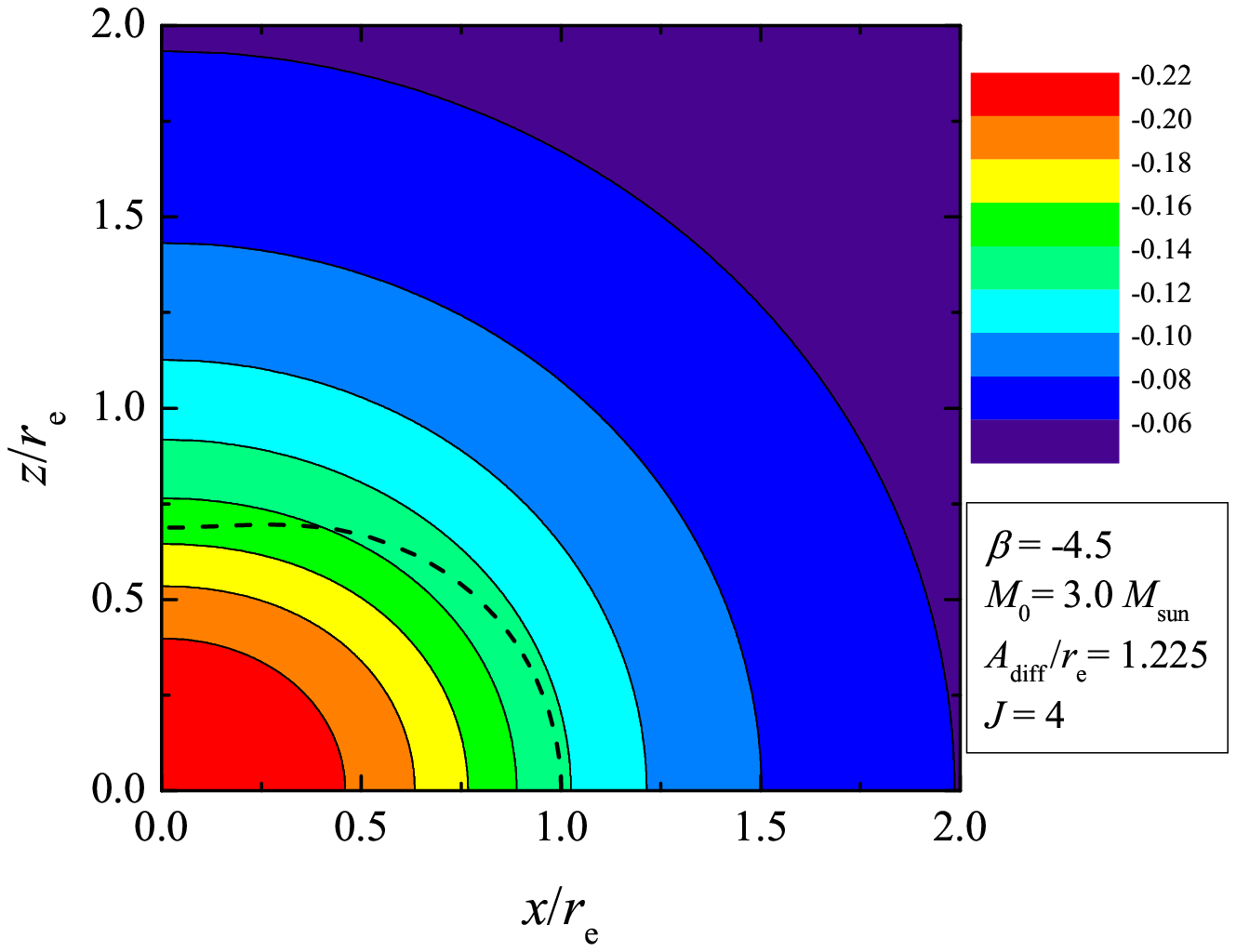}
	\includegraphics[width=0.45\textwidth]{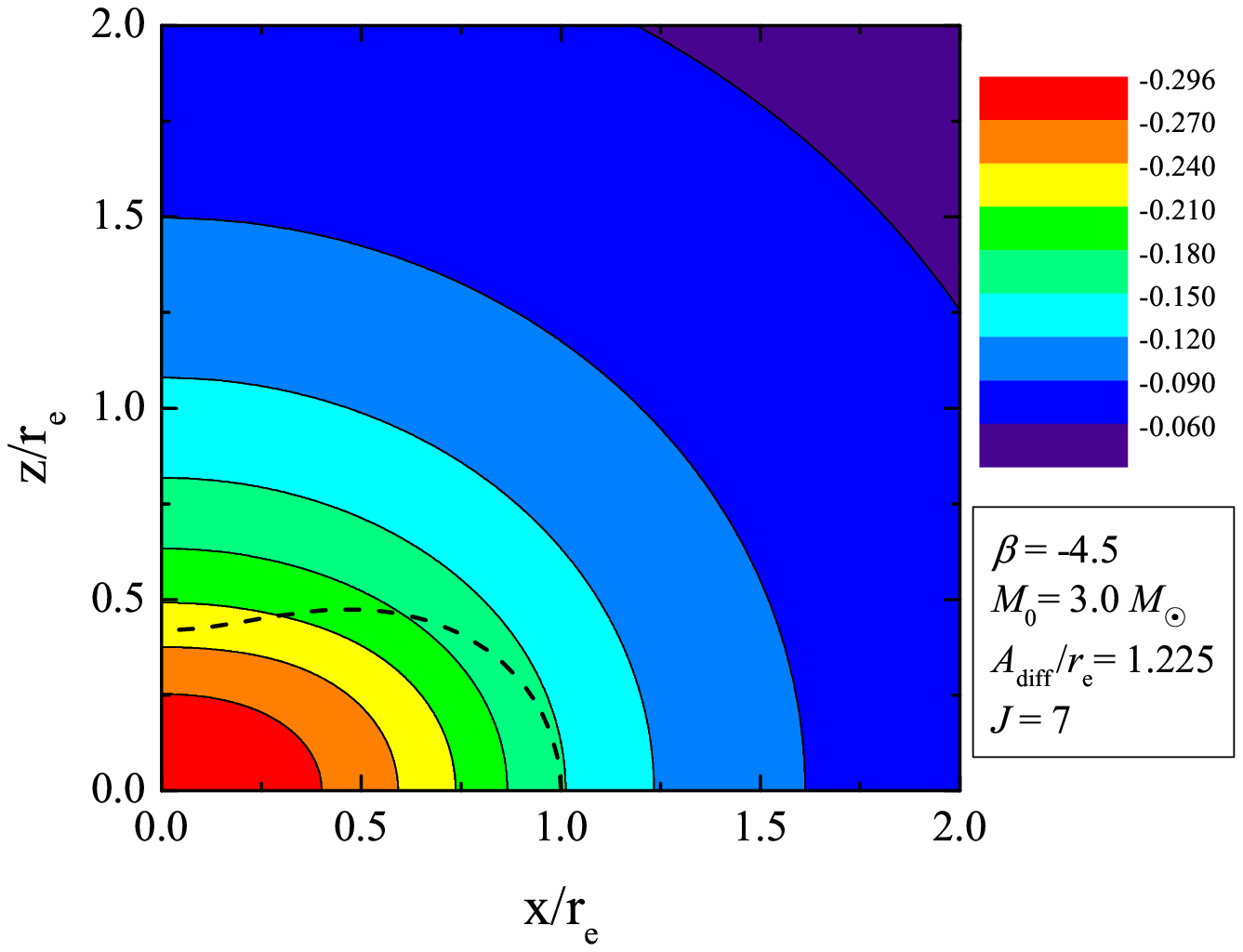}
	\caption{Contour plots of the scalar field. The two models are for fixed $\beta=-4.5$, ${\hat A} = A_{\rm diff}/r_e = 1.225$ and baryon mass $M_0=3.0$.  \textit{Left panel:}  $J=4$. \textit{Right panel:} $J=7$. }
	\label{Fig:Model_phi}
\end{figure}

\subsection{Structure of individual models}

In Fig. \ref{Fig:Model_J4eps}, contour plots of the neutron star energy density are presented for a model with baryon mass $M_b=3.0 M_\odot$ and angular momentum $J=4$, for both the GR case (left panel) and for STT with $\beta=-4.5$ (right panel). This specific value of the baryon mass is chosen because the merger of two standard mass neutron stars (say with gravitational mass approximately $1.3-1.4 M_\odot$) will result in a supramassive neutron star with baryon mass roughly $3.0 M_\odot$ for the APR4 EOS (the merger of two neutron stars is the most likely viable astrophysical scenario for creating neutron stars of such high masses). Two main differences between GR and STT are observed: a) for the same angular momentum, the STT model has a somewhat larger axis ratio and b) the core of the STT model is less quasi-toroidal than the core of the GR model. These differences should be attributed to the effect of the scalar field in the interior of the star.

In Fig. \ref{Fig:Model_J7eps}, we show a similar comparison as in Fig. \ref{Fig:Model_J4eps}, but for a higher angular momentum of $J=7$. 
This time, the two main differences between GR and STT (larger axis ratio and core which is less quasi-toroidal in STT than in GR) are much more pronounced than for the slower-rotating $J=4$ model.

Finally, Fig. \ref{Fig:Model_phi} shows the distribution of the scalar field (up to a distance of twice the coordinate equatorial radius) for the two models with $J=4$ and $J=7$. It is evident, that the scalar field is significantly less distorted away from spherical symmetry, when compared to the rotational distortion of the matter distribution. Even in the case of the rapidly rotating $J=7$ model, in which the matter distribution is highly quasi-toroidal, the scalar field remains quasi-spherical.

\section{Discussion}
\label{sec:conc}

We present the first numerical models of differentially rotating stars in alternative theories of gravity. Specifically, we studied models in 
 a particular class of scalar-tensor theories of gravity that is indistinguishable from GR in the weak field regime but can lead to significant deviations when strong fields are considered. The results are obtained using a new version of the {\tt RNS} code, employing the usual one-parameter rotation law for differential rotation. The degree of differential rotation considered in the paper is chosen so as to lead to bulk properties (mass, angular momentum and size) similar to those expected for binary neutron star merger remnants (see also \cite{Bauswein2017a}).

Sequences with fixed angular momentum $J$ are studied both in GR and in STT with different values of the coupling parameter $\beta$. We were able to construct constant $J$ sequences with relatively large values of the angular momentum and found that models in STT can reach values of  angular momentum that are not possible in GR. At the same time, models in STT can reach higher masses than it is possible in GR. Moreover, we find that scalarized solutions with sufficient angular momentum and values of $\beta$ that are in agreement with the observations in the massless STT case, possess a second turning-point along a constant-$J$ sequence, which appears at somewhat lower central density and is not present for smaller $J$ or in the nonrotating case. Since in GR turning points are associated with quasi-radial stability and the scalarized solution is energetically favoured over the GR solution (in the region where scalarized solutions appear) it will be important to study the dynamical properties of models along such constant-$J$ sequences, in order to determine whether an unstable region exists between the two turning points. The possible existence of such an unstable region would have astrophysical consequences for the stability of binary neutron star merger remnants.

We showed that the distribution of the scalar field retains a quasi-spherical shape, even for very rapidly rotating models, where the matter distribution has become strongly quasi-toroidal, even though that scalar field is directly sourced by the matter distribution in the right side of \eqref{EFFE}. The interaction of the scalar field with the matter through the hydrostationary equilibrium equation results in a matter distribution, which is less quasi-toroidal, when compared to a corresponding GR model with the same $J$. This, in turn, allows for the existence of models with higher $J$ and higher masses in STT than in GR.

A next step would be to consider a more realistic law for differential rotation, describing accurately the post-merger phase and to investigate the existence of universal relations, especially for the models which are at the threshold of collapse to a black hole (this threshold distinguishes prompt versus delayed collapse in binary neutron star mergers). Such universal relations may become useful in interpreting gravitational wave observations from neutron star mergers and allow for tests of alternative theories of gravity. Although we used a particular STT here, we expect our findings to be generic for many other alternative theories of gravity. We aim at addressing the above issues in forthcoming publications.
 
\acknowledgments{DD would like to thank the European Social Fund, the Ministry of Science, Research and the Arts Baden-W\"urttemberg for the support. DD is indebted to the Baden-W\"urttemberg Stiftung for the financial support of this research project by the Eliteprogramme for Postdocs. The support by the COST Actions  CA15117, CA16104, CA16214 and MP1304 is also gratefully acknowledged. This work was supported by the DAAD program  ``Hochschulpartnerschaften mit Griechenland 2016'' (Projekt 57340132). SY is supported partially by the Sofia University Grants No 80-10-73/2018 and No 3258/2017.}

\appendix

\bibliography{references}

\end{document}